%% file: main.ltx
\documentclass[letterpaper,twoside]{article}
\usepackage{natbib}
\usepackage{amsmath}
\usepackage{amsthm}
\usepackage{amscd}
\usepackage{amsfonts,amssymb}
\usepackage{url}
\def\urlprefix\url#1{\url{#1}} % fix silly BibTeX hyphenation
\providecommand{\url}[1]{\texttt{#1}}

\newtheorem{theorem}{Theorem}
\newtheorem{proposition}{Proposition}
\newtheorem{lemma}{Lemma}
\newtheorem{corollary}{Corollary}
\theoremstyle{definition}
\newtheorem{definition}{Definition}
\newtheorem{defiprop}{Definition and proposition}
\theoremstyle{remark}
\newtheorem{remark}{Remark}
\newtheorem{example}{Example}

\newcommand{\C}{{\mathbb{C}}}
\newcommand{\N}{{\mathbb{N}}}
\newcommand{\R}{{\mathbb{R}}}
\newcommand{\Z}{{\mathbb{Z}}}
\newcommand{\ONE}{{\boldsymbol{1}}}

\newcommand{\hookllongrightarrow}{\lhook\joinrel\relbar\joinrel\longrightarrow}
\def\<#1|#2>{\left\langle #1\vphantom{#2}\right|\left.\negmedspace\vphantom{#1}#2\right\rangle}

\DeclareMathOperator{\tr}{tr}
\DeclareMathOperator{\spec}{spec}
\DeclareMathOperator{\sgn}{sgn}
\DeclareMathOperator{\ord}{ord}
\DeclareMathOperator{\meas}{meas}

\begin{document}
\title{Measures of {F}ermi surfaces and absence of singular continuous spectrum for magnetic 
{S}chr{\"o}dinger operators}
\author{Michael J.\ Gruber\footnote{
Institut f\"ur Mathematik, Humboldt-Universit\"at zu Berlin, D-10099 Berlin, Germany.
\protect\url{mailto:gruber@mathematik.hu-berlin.de}
}
\footnote{
Department of Mathematics, MIT 2-167,
77 Massachusetts Avenue,
Cambridge, MA 02139-4307,
USA.
\protect\url{mailto:mjg@math.mit.edu}
}}
\maketitle
\begin{abstract}
\input{abstract}
\end{abstract}

\input{main}

%\bibliography{mrabbrev,publish,mathlib,gast,gruber,accepted,submitted}
%\bibliographystyle{gruberes}
\input main.bbl

\end{document}

%% file: abstract.tex
%auto-ignore
Fermi surfaces are basic objects in solid state physics and in the spectral theory of periodic operators.
We define several measures connected to Fermi surfaces and study their measure theoretic properties.
From this we get absence of singular continuous spectrum and of singular continuous components in the density of states for symmetric periodic elliptic differential operators acting on vector bundles.
This includes Schr\"odinger operators with periodic magnetic field and rational flux, as well as the corresponding Pauli and Dirac-type operators.

%% file: main.tex
%auto-ignore
\section*{Introduction}
Fermi surfaces are basic objects in solid state physics of metals.
They define the region in phase space which is relevant at room temperatures or below:
the only electrons which take part in transport are those differing by almost an energy $k_BT$ from the electrons on the Fermi surface which have energy $k_BT_F$.
Here, $k_B$ is Boltzmann's constant, $T_F$ is the Fermi temperature which is of the order of $10000\,K$ for ordinary metals compared to room temperatures of $300\,K$.
The curvature of the Fermi surface endows electrons with a dynamic mass, singularities of the Fermi surface can show up in measurements.

In the spectral theory of periodic elliptic operators, Bloch theory (also known as Floquet theory in the one-dimensional case) is the preferred tool. 
First of all it gives the band structure of the spectrum in a topological sense (locally finite union of closed intervals).
Many elements from solid state physics show up, although named differently.
In this paper we focus on the notion of Fermi surface and its importance for spectral theory.
Since it is defined by one real equation one expects it to be of codimension one in general (in a suitable sense) and therefore would have zero measure as a subset of the Brillouin zone.
On the other hand, for the 2-dimensional Schr\"odinger operator with constant magnetic field and zero potential (the Landau operator) the spectrum consists of a discrete set of infinitely degenerate eigenvalues, the Fermi 'surface' is either empty or given by the whole Brillouin zone (which has full measure, of course), depending on the Fermi energy.
This is definitely not considered as band structure by physicists, but it is band structure in the topological sense mentioned above.

One mathematical notion which distinguishes these situations is the measure theoretic type of the spectrum:  absolutely continuous, pure point, or singular continuous. 
Physically this corresponds to band structure, break-down of band structure, and physically not occuring structure (in our context of smooth periodic potentials and absence of disorder).
We define a (quasi-) measure which associates to each energy interval the measure of the corresponding Fermi shell in momentum space.
Its measure theoretic properties determine those of the spectrum.
The aim of this paper is to show that the physically unexpected (singular continuity) does not occur mathematically for symmetric periodic elliptic differential operators acting on sections of vector bundles over manifolds.
This class of operators contains the Schr\"odinger operator with periodic magnetic field and rational magnetic flux, the Pauli and the Dirac operator under the same conditions.

The spectral history of these operators is of course long. 
Related to our work is:
\citet{Tho:TDASIC} was the first to show absence of eigenvalues for periodic Schr\"odinger operators in $\R^n$. 
\citet{Ger:RTPSO} discussed the extension properties of resolvents for these operators, using analytic families of Fredholm determinants (we will use real-analytic families of $\zeta$-regularized determinants).
\citet{HemHer:BGPMH} attacked the periodic Sch\"odinger operator with constant magnetic field in $\R^n$ and showed absence of singular continuous spectrum in the case of zero flux (we will follow their ideas). 
They were able to prove absence of eigenvalues for small enough magnetic field also.
This was improved by
\citet{BirSus:TDPMHAC} who proved absence of eigenvalues in the 2-dimensional case with minimal regularity assumptions (operators defined by quadratic forms).
\citet{Sob:ACPMSO} proved the same in higher dimensions for more regular potentials.

In this paper we will not deal with singular potentials but assume all coefficients to be smooth.
This allows to work in a more geometric context which pays off already in the Euclidean case as soon as the magnetic flux is non-zero.

\subsection*{Outline}
In section \ref{sec:BTVB} we recall the basic notions of Bloch theory for symmetric periodic elliptic operators acting on sections of vector bundles.

In section \ref{sec:MFS} we define Fermi surfaces and construct several measures (Fermi measure, integrated density of states). We analyze their mutual relations and their relations to the spectrum.

In section \ref{sec:SC} we investigate the properties of the Fermi measure and draw our conclusions for the spectrum.

\subsection*{Acknowledgements} 

This work is a (commutative) part of my Ph.D.\ thesis ``Nichtkommutative Blochtheorie'' \cite[non-commutative Bloch theory;][]{Gru:NB}. 
I gratefully appreciate the advice and supervision given by Jochen Br\"uning at Humboldt-University at Berlin. 
\nocite{Gru:NBTO}

This work was supported by Deutsche Forschungsgemeinschaft as project D6 at the Sonderforschungsbereich 288 (differential geometry and quantum physics).

%%%%%%%%%%%%%%%%%%%%%%%%%%%%%%%%%%%%%%%%%%%%%%%%%%%%%%%%%%%%%%%%%%%%%%%%%%%%%%%%%%%%%%%%%%%%%
\section{Bloch theory on vector bundles}\label{sec:BTVB}

In this section we recall the basic elements of Bloch theory for periodic operators in the geometric context of  vector bundles, since even in the scalar case of a magnetic Schr\"odinger operator
one is lead to consider
 possibly non-trivial complex line bundles.
The standard reference for the theory of direct integrals is \citep[chapter II]{Dix:AOEHAN}, for Bloch theory in Euclidean space see \citep[chapter XIII.16]{ReeSim:AO}. 

Our general assumptions are: $X$ is an oriented smooth Riemannian manifold without boundary, $\Gamma$ a discrete abelian group acting on $X$ freely, isometrically, and properly discontinuously.
Furthermore, we assume the action to be cocompact in the sense that the quotient $M:=X/\Gamma$ is compact.

Next, let $E$ be a smooth Hermitian vector bundle over $X$.

\begin{example}[solid crystals]
The main motivating example for our setting comes from solid state physics.
Here, $X=\R^n$ is the configuration space of a single electron ($n=2,3$). 
It is supposed to move in a crystal whose translational symmetries are described by a lattice $\Z^n\simeq\Gamma\subset\R^n$, which acts on $X$ by translations, of course.
Note that this does not take into account the point symmetries. 
$\Gamma$ could be extended by them but the action would not be free any more.
Considering just the translations is enough to achieve the compactness of the quotient $M\simeq T^n.$

Wave functions of electrons are just complex-valued functions on $\R^n$, so we can set $E=\R^n\times\C$.
One may also include the spin of the electrons into the picture by choosing the appropriate trivial spinor bundle $E=\R^n\times\C^2$.
\end{example}

\begin{definition}[periodic operator]\label{defi:periodic operator}
Assume there is an isometric  lift $\gamma_*$ of the action of $\gamma$ fom $X$ to $E$ in the following sense:
\begin{align}
\gamma_*:E_x\rightarrow E_{\gamma x}\text{ for }x\in X,\gamma\in\Gamma.
\end{align}
This defines an action $T_\gamma$ on the sections: For $s\in C^\infty_c(E)$ we define
\begin{align}
(T_\gamma s)(x) := \gamma_* s(\gamma^{-1}x)\text{ for }x\in X,\gamma\in\Gamma. \label{equ:actiononsections}
\end{align}
$(T_\gamma)_{\gamma\in\Gamma}$ induces a unitary representation of $\Gamma$ in $L^2(E)$ since $\gamma_*$ acts isometrically and $T_\gamma^*=(T_\gamma)^{-1}$.

A differential operator $D$ on ${\mathcal D}(D):=C^\infty_c(E)$ is called periodic if, on $\mathcal{D}(D)$, we have: 
\begin{equation} \forall{\gamma\in\Gamma}: [T_\gamma,D]=0 \end{equation} 
\end{definition}

\begin{example}[periodic Schr\"odinger operator]
Given a manifold as described above, we may lift the action to any trivial vector bundle $E:=X\times\C^k$ canonically.
If $D$ is a periodic operator on $X$ (for example any geometric operator, i.e.\ defined by the metric on $X$) and $V\in C^\infty(X,M(k,\C))$ a periodic field of endomorphisms, then $D+V$ is a periodic operator on $E$.

In the case of a crystal, we choose the Laplacian (which describes the kinetic energy quantum mechanically) and a periodic potential $V\in C^\infty(\R^n,\R)$ (which describes the electric field of the ions at the lattice sites) to get the periodic Schr\"odinger operator $\Delta+V$. 
\end{example}

\begin{example}[Schr\"odinger operator with exact periodic magnetic field]
Let $b\in\Omega^2(X)$ be a magnetic field 2-form. 
In dimension 3 this corresponds (by the Hodge star) to a vector field $B$, in dimension 2 to a scalar function which may be thought of as the length and orientation of a normal vector $B$.
From physical reasons one has $\operatorname{div} B=0$, i.e.\ $db=0$.
For simplicity we assume that $b$ is not only closed but exact, so there is $a\in\Omega^1(X)$ with $b=da$ ($B=\operatorname{rot} A$ for the corresponding vector fields).
This defines a magnetic Hamiltonian operator
\begin{align} \Delta^a:=(d-\imath a)^*(d-\imath a) \label{equ:BL} \end{align}
(the \emph{minimally coupled Hamiltonian}), where $d$ is the ordinary differential (corresponding to the gradient) and ${}^*$ the  adjoint of an operator between the Hilbert spaces of $L^2$-functions $L^2(X)$ and of $L^2$-1-forms $L^2(X,\Lambda T^*X)$.

\def\ma{{\underline{*}}}
For later convenience we set, for $\gamma\in\Gamma$ and $\omega\in\Omega(X)$, $\gamma^\ma\omega:=\left(\gamma^{-1}\right)^*\omega$, considering $\gamma^{-1}$ as a map $X\rightarrow X$ and using the usual pull-back of forms. This puts the action on forms in a notation compatible with the action on sections~\eqref{equ:actiononsections} from the preceding definition.

Now, if $b$ is periodic, $a$ does not need to be so: If $b\in\Omega^2(\R^n)$ is constant then $a$ is affine linear. So the translations are no symmetries for the magnetic Hamiltonian. 
\citet{Zak:DESEF} was the first to define the so-called \emph{magnetic translations}: Since $d(a-\gamma^\ma a)=da-\gamma^\ma da=b-\gamma^\ma b=0$, one can (at least if $H^1(X)=0$) find a function $\chi_\gamma$ with $d\chi_\gamma=a-\gamma^\ma a$. 
One may define such a function  explicitly by
\[ \chi_\gamma(x):=\int_{x_0}^x(a-\gamma^\ma a) \]
which is well-defined if $H_1(X)=0$.
If we now define a gauge function $s_\gamma := e^{\imath\chi_\gamma}$ then
\begin{align*}
  (d-\imath a)(s_\gamma \gamma^\ma f) &= s_\gamma \gamma^\ma df + \imath(a-\gamma^\ma a)s_\gamma 
\gamma^\ma f -\imath a s_\gamma \gamma^\ma f \\
 &= s_\gamma \gamma^\ma df -\imath \gamma^\ma a\gamma^\ma f \\
 &= s_\gamma \gamma^\ma df -\imath s_\gamma \gamma^\ma(a f) \\
 &= s_\gamma \gamma^\ma \left( (d-\imath a) f \right).
\end{align*}
So we have found symmetries of the magnetic Hamiltonian operator, the gauged translations
\begin{align*}
T_\gamma: C^\infty(X) &\rightarrow C^\infty(X), \\
( T_\gamma s)(x) &= s_\gamma(x) (\gamma^\ma s)(x)
\end{align*}
coming from the lifted action
\begin{align*}
\gamma_*: X\times\C&\rightarrow X\times\C,\\
 \gamma_*(x,c) &= \left(\gamma x, s_\gamma(x)c\right).
\end{align*}
The commutation relation for the magnetic translations is
\begin{align}
(T_{\gamma_1}T_{\gamma_2}s)(x) &= s_{\gamma_1}(x)s_{\gamma_2}(\gamma_1^{-1}x)s(\gamma_2^{-1}\gamma_1^{-1}x) \notag \\
&=\exp\left(\imath\left(\int_{x_0}^x a-\gamma_1^\ma a+\int_{x_0}^{\gamma_1^{-1}x}a-\gamma_2^\ma a \right)\right) s(\gamma_2^{-1}\gamma_1^{-1}x) \notag \\
&=\exp\left(\imath\left(\int_{x_0}^x a-\gamma_1^\ma a+\int_{\gamma_1x_0}^{x}\gamma_1^\ma a-(\gamma_1\gamma_2)^\ma a \right)\right) s(\gamma_2^{-1}\gamma_1^{-1}x) \notag   \\
&=\exp\left(\imath\left(\int_{x_0}^{\gamma_1x_0} (\gamma_1\gamma_2)^\ma a-\gamma_1^\ma a + \int_{x_0}^x a-(\gamma_1\gamma_2)^\ma a\right)\right) s(\gamma_2^{-1}\gamma_1^{-1}x)\notag  \\
&=\exp\left(\imath\left(\int_{x_0}^{\gamma_1x_0} (\gamma_1\gamma_2)^\ma a-\gamma_1^\ma a\right)\right) s_{\gamma_1\gamma_2}(x)s(\gamma_2^{-1}\gamma_1^{-1}x) \label{equ:flux} \\
&=:\Theta(\gamma_1,\gamma_2)s_{\gamma_1\gamma_2}(x)s(\gamma_2^{-1}\gamma_1^{-1}x)\notag \\
&=\Theta(\gamma_1,\gamma_2)(T_{\gamma_1\gamma_2}s)(x) \notag
\end{align}
with $\Theta(\gamma_1,\gamma_2)\in S^1$. In general this is just a projective representation of $\Gamma$.
If $a$ itself  is periodic, then $\chi_\gamma=0$ for $\gamma\in\Gamma$, i.e.\ there is no gauge, and we have just ordinary translations forming a proper representation.

But even if $a$ is not periodic it can happen that the magnetic translations commute with each other. This is called the case of \emph{integral flux} since the term occuring in the exponential in line~\eqref{equ:flux} is just the magnetic flux through one lattice face. A periodic $a$ obviously gives rise to zero magnetic flux.

Furthermore, if $V\in C^\infty(X,\R)$ is $\Gamma$-periodic it commutes with the magnetic translations as well, so $\Delta^a+V$ is a (symmetric elliptic) periodic operator.

Finally, the very same magnetic translations can be used for the Pauli Hamiltonian and the magnetic Dirac operator.
\end{example}

\begin{remark}[non-integral flux]
If the magnetic flux is rational one can find a superlattice of $\Gamma$, i.e.\ a subgroup of finite index, such that the flux is integral. The quotient will still be compact, of course, so that the rational case can be completely reduced to the integral.

If the magnetic flux is irrational there is no such superlattice. 
Still, one may try to make use of the projective representation defined above.
There are several approaches, similar in the objects which are used, different in the objectives that are aimed at and accordingly in the results. 
For references and one approach which mimicks Bloch theory non-commutatively we refer to \citep{Gru:NBTO}.
\end{remark}

\begin{remark}[non-exact magnetic field]
If $b$ is closed but not exact one first has to agree upon the quantization procedure used. 
\eqref{equ:BL} may be identified as a Bochner Laplacian for a connection with curvature $b$, and such a connection exists if and only if $b$ defines an integral cohomology class, i.e.\ $[b]\in H^2(X,\Z)$.
There may exist different quantizations for the same magnetic field.
This is connected to the Bloch decomposition again, For this and the construction of the magnetic translations in this case see \citep{Gru:BTQMS}.
\end{remark}

\begin{lemma}[associated bundle]
$E$ is the lift  $\pi^*E'$ of a Hermitian vector bundle $E'$ over $M$ by the projection $\pi:X\rightarrow M$. $E$ and $X$ are $\Gamma$-principal fiber bundles over $E'$ resp.\ $M$.

To every $\Gamma$-principal fiber bundle and every character $\chi\in\hat\Gamma$
we associate a line bundle. 
This gives the relations depicted in the following diagram
  (``$\rightsquigarrow$'' denotes association of line bundles.):

%\begin{figure}[htbp] \begin{minipage}{\textwidth}
\begin{equation*}
\begin{CD}
       @.   \C^N @.   \C^N  @.  @. @. \C^N @. \C^N\\
   @.       @VVV      @VVV   @. @. @VVV @VVV  \\
\Gamma &\hookllongrightarrow& E    @>\textstyle\pi_*>> E'    @. \quad\rightsquigarrow\quad @. \C &\hookllongrightarrow& E_\chi    @>>> E' \\
   @.       @VV\textstyle\pi^EV      @VV\textstyle\pi^{E'}V @.                          @. @VVV @VVV \\
\Gamma &\hookllongrightarrow& X    @>\textstyle\pi>> M   @. \quad\rightsquigarrow\quad @. \C &\hookllongrightarrow& F_\chi    @>>> M 
\end{CD}
\end{equation*}
%\caption{
\begin{center}principal fiber bundles and associated line bundles\end{center}\label{fig:PAV}
%\end{minipage}
%\end{figure}

In this situation we have $E_\chi\simeq E'\otimes F_\chi$.
\end{lemma}
\begin{proof}
$E$ is a $\Gamma$-principal fiber bundle, so we can use the lifted  $\Gamma$-action  to define $E':=E/\Gamma$.
Since this action is a lift of the $\Gamma$-action on $X$, $E'$ has a natural structure of a vector bundle over $M$. If $\pi^{E'}:E'\rightarrow M$ is the bundle projection of $E'$, then the pull back by $\pi$  is defined as
\begin{align*}
\pi^* E' &= X\times_\pi E'=\{ (x,e)\in X\times E'\mid \pi(x)=\pi^{E'}(e) \}.
\end{align*}
If $\pi^{E}:E\rightarrow X$ is the bundle projection of $E$ and $\pi_*:E\rightarrow E'$ is the quotient map, then we get a bundle isomorphism $E\rightarrow \pi^* E'$  by
\begin{align*}
E\ni e \mapsto (\pi^E(e),\pi_*(e)) \in \pi^* E'.
\end{align*}
Therefore, in this representation the lift $\gamma_*$ of $\gamma$ acts on $(x,e)\in\pi^*E'$ as $\gamma_*(x,e)=(\gamma x,e)$.

Sections into an associated bundle $P\times_\rho V$ are just those sections of the bundle $P\times V$ which have the appropriate transormation property. 
By construction, $E_\chi$ is a complex line bundle over $E'$, but from $E$ it inherits the vector bundle structure, so its sections fulfill:
\begin{equation} C^\infty(E_\chi) \simeq C^\infty(E)^{\Gamma,\chi}=\{s\in C^\infty(E)\mid \forall{\gamma\in\Gamma}:\gamma^*s=\chi(\gamma)s \} \label{equ:Echi=EGchi} \end{equation}
An analogous equation holds for the line bundle $F_\chi$ over $M$.
Finally, \eqref{equ:Echi=EGchi} shows
\begin{align*} E_\chi &= E\times_\chi\C \\
 &= (\pi^* E')\times_\chi\C \\
 &= (X\times_\pi E')\times_\chi\C \\
 &\simeq E'\otimes (X\times_\chi\C) \\
 &= E'\otimes F_\chi.
\end{align*}
Here, all equalities are immediate from the definitions, besides the last but one, which may be seen as follows:
\begin{align*}
(X\times_\pi E')\times_\chi\C &= (X\times_\pi E'\times\C)/\Gamma
\intertext{with the $\Gamma$-action}
\gamma(x,e,z) &= (\gamma x,e,\chi(\gamma)z),
\intertext{whereas}
E'\otimes (X\times_\chi\C) &=E'\otimes ((X\times\C)/\Gamma)
\intertext{with the $\Gamma$-action}
\gamma(x,z) &= (\gamma x,\chi(\gamma) z).
\end{align*}
So, both bundles are quotients of isomorphic bundles with respect to the same $\Gamma$-action.
\qed \end{proof}

\begin{example}[magnetic bundles]
Consider again the case of the magnetic translations for a periodic magnetic 2-form $b\in\Omega^2(X)$, $E$ being a complex line bundle with curvature $b$ ($b\in H^2(X,\Z)$). Hence we have $c_1(E)=[b]$ for the Chern class (up to factors of $2\pi$, depending on the convention).
Since $b$ is periodic we may restrict it to a form $b_M\in\Omega^2(M)$ on the quotient.
The existence of the lifted action, i.e.\ the fact that $E$ can be written as a pull-back $E=\pi^*E'$, corresponds to the integrality of $b_M$ from $c_1(E')=[b_M]\in H^2(M,\Z)$. 
Tensoring $E'$ with the flat line bundle $F_\chi$ does not change the Chern class (up to torsion).
In particular, in dimension 2 the integrality of $b_M$ is equivalent to the integrality of the flux, and $E'$ is trivial only for zero flux. 
\end{example}

Next we want to decompose the Hilbert space $L^2(E)$ of square-integrable sections of $E$ into a direct integral over the character space $\hat\Gamma$. 
On $\hat\Gamma$ we use the Haar measure. 
From the theory of representations of locally compact groups we need the following character relations for abelian discrete $\Gamma$, i.e.\ for abelian, compact $\hat\Gamma$ \cite[see e.g.\ ][ \S 1.5]{Rud:FAG}:

\begin{lemma}[character relations]
For $\gamma\in\Gamma$ 
\begin{equation} \int_{\hat\Gamma}\chi(\gamma)\,d\chi = \left\{ \begin{array}{l} 1, \quad \gamma=e, \\ 0, \quad \gamma\neq e. \end{array} \right. \label{equ:Charakter}\end{equation}
For $\chi,\chi'\in\hat\Gamma$ 
\begin{equation} \sum_{\gamma\in\Gamma} \bar\chi(\gamma)\chi'(\gamma) = \delta(\chi-\chi') \label{equ:Charakter'}\end{equation}
in distributional sense, i.e.\ for $f\in C(\hat\Gamma)$ 
\[ \sum_{\gamma\in\Gamma} \int_{\hat\Gamma}\bar\chi(\gamma)\chi'(\gamma)f(\chi)\,d\chi = f(\chi'). \]
\end{lemma}

We define for every character $\chi\in\hat\Gamma$ a mapping $\Phi_\chi:C^\infty_c(E)\ni s\mapsto \tilde s_\chi\in C^\infty(E)$ by
\begin{equation} \tilde s_\chi(x) := \sum_{\gamma\in\Gamma} \chi(\gamma)\gamma_*s(\gamma^{-1}x). \label{equ:schi} \end{equation}
Since
\begin{align*}
 \tilde s_\chi(\gamma' x) &= \sum_{\gamma\in\Gamma}\chi(\gamma)\gamma_*s(\gamma^{-1}\gamma'x) \\
 &= \sum_{\gamma\in\Gamma}\chi(\gamma'\gamma'{}^{-1}\gamma)(\gamma'\gamma'{}^{-1}\gamma)_*s\left((\gamma'{}^{-1}\gamma)^{-1}x\right) \\
 &= \chi(\gamma')\gamma'_* \tilde s_\chi(x)
\end{align*}
we have
\[ \tilde s_\chi\in C^\infty(E)^{\Gamma,\chi}=\{r\in C^\infty(E)\mid \forall_{\gamma\in\Gamma} T_\gamma r = \chi(\gamma) r\} \] 
which defines a section $s_\chi\in C^\infty(E_\chi)$. 

Let $\mathcal D$ be a fundamental domain for the $\Gamma$-action, i.e.\ an open subset of $X$ such that $\bigcup_{\gamma\in\Gamma}{\gamma\mathcal D}=X$ up to a set of measure 0 and $\gamma\mathcal D\cap\mathcal D=\emptyset$ for $\gamma\not=e$. Then
\begin{align*}
 \int_{\hat\Gamma} \|s_\chi\|^2_{L^2(E_\chi)}d\chi &= \int_{\hat\Gamma} \int_{\mathcal{D}} |\tilde s_\chi(x) |^2 dx\,d\chi \\
 &= \int_{\mathcal{D}} \int_{\hat\Gamma} \sum_{\gamma_1,\gamma_2\in\Gamma} \chi(\gamma_1^{-1}\gamma_2)\langle {\gamma_1}_*s(\gamma_1^{-1}x)\mid{\gamma_2}_*s(\gamma_2^{-1}x)\rangle_{E} d\chi\, dx \\
 &= \int_{\mathcal{D}} \sum_{\gamma\in\Gamma} |s(\gamma^{-1}x)|^2 dx \\
 &= \|s\|^2_{L^2(E)}.
\end{align*}
On the one hand, this shows that we can define a measurable structure on $\prod_{\chi\in\hat\Gamma}L^2(E_\chi)$ by choosing a sequence in $C^\infty_c(E)$ which is total in  $L^2(E)$.
On the other hand, we can see that the direct integral $\int^\oplus_{\hat\Gamma}L^2(E_\chi)\,d\chi$ is isomorphic to $L^2(E)$ via the isometry
$\Phi$, whose inverse is given by
\[ \Phi^*\colon(s_\chi)_{\chi\in\hat\Gamma} \mapsto \int_{\hat\Gamma} \tilde s_\chi(x)\,d\chi, \]
as is easily seen from the character relations \eqref{equ:Charakter} and \eqref{equ:Charakter'}.

This shows

\begin{lemma}[direct integral]
 The mapping defined by \eqref{equ:schi} can be extented continuously to a unitary
\begin{equation} \Phi: L^2(E) \rightarrow \int^\oplus_{\hat\Gamma} L^2(E_\chi)\,d\chi. \end{equation}
\end{lemma}

For the direct integral of Hilbert spaces $H=\int^\oplus_{\hat\Gamma}H_\chi d\chi$ 
the set of decomposable bounded operators
$L^\infty(\hat\Gamma,{\mathcal L}(H))$  is given by the commutant
$(L^\infty(\hat\Gamma,\C))'$ in ${\mathcal L}(H)$.
Since commutants are weakly closed and $C(\hat\Gamma,\C)$ is weakly dense in $L^\infty(\hat\Gamma,\C)$ one has $(L^\infty(\hat\Gamma,\C))'=(C(\hat\Gamma,\C))'$.
Therefore, in order to determine the decomposable  operators one has to determine the action of $C(\hat\Gamma)$ on $L^2(E)$. This is easily done using the explicit form of $\Phi$:

\begin{proposition}[$C(\protect\hat{\Gamma})$-action]
$f\in C(\hat\Gamma)$ acts on $s\in C^\infty_c(E)$ by
\begin{align}
 M_fs &:= \Phi^*f\Phi s, \\
\intertext{and one has}
 (M_fs)(x) &= \sum_{\gamma\in\Gamma} \hat f(\gamma^{-1})T_\gamma s(x),\text{ where} \label{equ:fAktion} \\
 \hat f(\gamma) &:= \int_{\hat\Gamma}f(\chi)\bar\chi(\gamma)\,d\chi \label{equ:fhat=intf}
\end{align}
is the Fourier transform of $f$. 
$M_f$ is a bounded operator with norm $\|f\|_\infty$.
\end{proposition}
\begin{proof}
For $x\in X$ one has:
\begin{align*}
 (M_fs)(x) &= (\Phi^*f\Phi s)(x) \\
 &= \int_{\hat\Gamma} (f\Phi s)_\chi(x)\,d\chi \\
 &= \int_{\hat\Gamma} f(\chi) \sum_{\gamma\in\Gamma}\chi(\gamma)\gamma_* s(\gamma^{-1}x)\,d\chi \\
 &= \sum_{\gamma\in\Gamma} \hat f(\gamma^{-1})\gamma_* s(\gamma^{-1}x)
\end{align*} 
Since $f$ is a multiplication operator in each fiber it has fiberwise norm $\|f\|_\infty$, and so have $f$ and $M_f=\Phi^*f\Phi$.
\qed \end{proof}

\begin{corollary}[decomposable operators]
Conjugation by $\Phi$ defines an isomorphism between decomposable bounded operators on $\int^\oplus_{\hat\Gamma}L^2(E_\chi)\,d\chi$ and  $\Gamma$-periodic bounded operators on $L^2(E)$.
\end{corollary}
\begin{proof} \item[]
\begin{description}
\item[``$\Rightarrow$''] A decomposable operator commutes with the $C(\hat\Gamma)$-action, especially with $f_\gamma\in C(\hat\Gamma)$ which is defined by
\[ \hat f_\gamma(\gamma') := \begin{cases} 1,& \text{if }\gamma=\gamma', \\ 0& \text{else.}  \end{cases} \]
By~\eqref{equ:fAktion} commuting with $f_\gamma$ is equivalent to commuting with $\gamma$.
\item[``$\Leftarrow$'']  To commute with the $\Gamma$-action means to commute with all $f_\gamma$ for $\gamma\in\Gamma$. Because of
\[ f_\gamma(\chi) = \chi(\gamma) \]
the $f_\gamma$ are just the characters $\widehat{\hat\Gamma}$ of the compact group $\hat\Gamma$, and by the Peter-Weyl theorem (or simpler: by the Stone-Weierstra\ss{} theorem) they are dense in $C(\hat\Gamma)$. 
Since the operator norm  of $M_f$ and the  supremum norm of $f$ coincide the commutation relation follows for all $f\in C(\hat\Gamma)$ by continuity.
\qed  \end{description}
\end{proof}

An unbounded operator is decomposable if and only if its (bounded) resolvent is decomposable.
For a periodic symmetric elliptic operator $D$ we have a domain of definition ${\mathcal D}(D)=C^\infty_c(X)$ on which $D$ is essentially self-adjoint.
This domain is invariant for $D$ as well as for the $\Gamma$-action,
and  one has $[D,\gamma]=0$ for all $\gamma\in\Gamma$. 
Thus all bounded functions of $D$ commute with the $\Gamma$-action, and one has:

\begin{theorem}[decomposition of periodic operators]\label{theorem:DPO}
The closure $\bar D$ of every periodic symmetric elliptic operator $D$ is decomposable with respect to the direct integral $\int^\oplus_{\hat\Gamma}L^2(E_\chi)\,d\chi$.
A core for the domain of $\bar D_\chi$ is given by $C^\infty(E_\chi)$, and the action of  $D_\chi$ on $C^\infty(E_\chi)\simeq C^\infty(E)^{\Gamma,\chi}$ is just the action of $D$ as differential operator on $C^\infty(E)^{\Gamma,\chi}$. 
We have $\bar D_\chi=\overline{D_\chi}$, where 
\begin{equation} D_\chi:=D|_{C^\infty(E)^{\Gamma,\chi}} \label{equ:Dchi=D|Echi}\end{equation}
 and the closures are to be taken as operators in $L^2(E_\chi)$.
\end{theorem}
\begin{proof}
Given the remark above we have shown the decomposability already.

 $C^\infty_c(X)$ is a core for $\bar D$, its image under $\Phi_\chi$ is contained in $C^\infty(E)^{\Gamma,\chi}$ and is a core for $\bar D_\chi$, sinece $\Phi$ is an isometry.
On this domain  \eqref{equ:schi} gives the action of $\bar D_\chi$ as asserted in the theorem.
Since $D_\chi$ is a symmetric elliptic operator on the compact manifold $M$ it is essentially self-adjoint. $\bar D_\chi$ is a fiber of $\bar D$ \citep[which is self-adjoint by, e.g.,][]{Ati:EODGNA} and therefore self-adjoint, thus both define the same unique self-adjoint extension $\overline{D_\chi}$ of $D_\chi$. 
\qed \end{proof}

In passing we harvest a corollary which we will not use in the sequel, but which is well known in the Euclidean setting:
\begin{corollary}[reverse Bloch property]\label{corol:RBP}
Every symmetric elliptic abelian periodic operator has the reverse Bloch property,  i.e.\ to every $\lambda\in\spec \bar D$ there is a bounded generalized eigensection $s\in C^\infty(E)$ with $Ds=\lambda s.$
\end{corollary}
\begin{proof}
If $\lambda\in\spec \bar D$ then, by the general theory for direct integrals, 
\[ \{\chi\in\hat\Gamma\mid (\lambda-\varepsilon,\lambda+\varepsilon)\cap\spec \bar D_\chi\neq\emptyset\} \]
has positive measure for every $\varepsilon>0$. The fibers $\bar D_\chi$ are elliptic operators on a compact manifold and thus have  discrete spectrum;
the eigenvalues depend continuously on $\chi$ (even piecewise real-analytically; see below).
We choose a sequence $(\chi_n)_{n\in\N}$ with $(\lambda-1/n,\lambda+1/n)\cap\spec \bar D_{\chi_n}\neq\emptyset$, so that there is an accumulation point $\chi_\infty$ ($\hat\Gamma$ is compact), and  $\lambda\in\spec\bar D_{\chi_\infty}$ due to continuity.

Since $\spec\bar D_{\chi_\infty}$ is discrete $\lambda$ is an eigenvalue of $\bar D_{\chi_\infty}$. 
The lift of an eigensection (which is smooth due to ellipticity) lies in $C^\infty(E)^{\Gamma,\chi}$ and therefore is bounded. 
Furthermore the lift satisfies the same eigenvalue equation because of \eqref{equ:Dchi=D|Echi}.
\qed \end{proof}

%%%%%%%%%%%%%%%%%%%%%%%%%%%%%%%
\section{Measures of Fermi surfaces}\label{sec:MFS}
In the previous section we noticed that the spectrum of an abelian-periodic operator $D$ can be computed form the spectra of its Bloch fibers which have discrete, finitely degenerate spectra.
In this section we will assume that $D$ is bounded from below since certain tools which we use can only be defined in this case.
In our main theorem about the spectrum (section~ \ref{sec:SC}, theorem~\ref{theorem:SNAPO}) we will show how to reduce the general case to the semi-bounded.

\begin{definition}[Fermi surface]
Let $D$ be an abelian-periodic symmetric operator, $D_\chi$ its Bloch fibers, and
 $E_0(\chi)\leq E_1(\chi)\leq E_2(\chi)\ldots$ the corresponding eigenvalues repeated according to multiplicity.
 \begin{align}
E_n:\hat\Gamma\rightarrow\R,\chi\mapsto E_n(\chi)\end{align} defines the $n$-th \emph{energy band}  (in the ``
reduced zone scheme'').
The  \emph{Fermi surface of the $n$-th band at energy $E$} is defined to be \begin{equation} F^D_n(E):=\{\chi\in
\hat\Gamma\mid E_n(\chi)=E\}, \end{equation}
the \emph{Fermi surface at energy $E$} is \begin{align}F^D(E)&:= \bigcup_{n\in\N_0}F^D_n(E)
\\ &\phantom{:}= \{\chi\in\hat\Gamma\mid E\in\spec D_\chi \}. \notag \end{align}
\end{definition}

``Usually'' the Fermi surface is of codimension 1 (if this notion makes sense) so that its measure vanishes, whereas
 $E$ is an eigenvalue of $D$ if and only if $\meas F^D(E)>0$ (from the direct integral).

\begin{defiprop}[Fermi measure]
For every Borel set $B\subset\R$ of energies we define the measure of its Fermi shell by
\begin{align} \mu^{F(D)}(B) &:= \meas \bigcup_{E\in B}F^D(E) \\
 &\phantom{:}= \meas \{ \chi\in\hat\Gamma\mid \spec D_\chi\cap B\ne\emptyset \} \notag
\end{align}
This is a quasi-measure (i.e.\ it is a measure up to additivity which is replaced by sub-additivity).
$E$ is an atom of $\mu^{F(D)}$ if and only if $\mu^{F(D)}(\{E\})>0$ if and only if $E$ is an eigenvalue of $D$.
\end{defiprop}
\begin{proof}
First, the energy band functions are measurable since they are the eigenvalue family of a decomposable operator in a direct integral; so the Fermi surface and Fermi shell are measurable.
Since the band functions are continuous and $\hat\Gamma$ is bounded, the measure is finite on bounded sets.

The statement about the atoms follows from the remark above, referring to the basic properties of direct integrals, since by definition $\mu^{d_D}(B)=\meas F^D(E)$.

If $E_1$ and $E_2$ are different it may happen that $F^D(E_1)$ and $F^D(E_2)$ intersect, but still
\[ \meas  \left( F^D(E_1) \cup F^D(E_2) \right) \leq \meas   F^D(E_1) + \meas F^D(E_2) \]
and analogous for disjoint Borel sets, so that $\mu^{F(D)}$ is sub-additive.
\qed \end{proof}

\begin{definition}[integrated density of states]
Let $P^D(E)$ be the spectral projections of $D$.
 The \emph{integrated density of states} (IDS) of $D$ is defined by
\begin{align} \mathcal{N}(E)&:=\tr_\Gamma P^D(E) \\
 &:=\int_{\hat\Gamma} \tr \left(P^D(E)\right)_\chi\,d\chi=\int_{\hat\Gamma} \tr P^{D_\chi}(E)\,
d\chi \notag
\end{align}
The IDS defines a Lebesgue--Stieltjes measure $\mu^{\mathcal{N}}$ such that $\mu^{\mathcal{N}}((-\infty,E])=\mathcal{N}(E)$.
\end{definition}

For $f\in L^2(E)$ we have the spectral measure at $f$ given by
\begin{align}
\mu_f^D(B) &=\int 1_B\,d\<f|P^D(E)f> \\
 &= \<f|P_B^Df> \\
 &= \int_{\hat\Gamma} \mu_{f_\chi}^{D_\chi}\,d\chi
\end{align}
for Borel sets $B\subset\R$. This is the usual definition from spectral theory.

\begin{definition}[determinant measure]
$(D_\chi)_{\chi\in\hat\Gamma}$ is a family of semi-bounded elliptic differential operators.
Then we have the family $\left(d_D(\cdot,\chi)\right)_{\chi\in\hat\Gamma}$ of $\zeta$-regularized determinants $d_D(\lambda,\chi)=\operatorname{det} (D_\chi -\lambda)$. 
The {associated quasi-measure} is defined by
\begin{equation}\mu^{d_D}(B):=\meas\{\chi\in\hat\Gamma\mid\exists\lambda\in B:d_D(\lambda,\chi)
=0\}\end{equation}
for every Borel set $B\subset\R$.
\end{definition}
The fact that $\mu^{d_D}$ is a quasi-measure can be seen as for the Fermi measure. 
We will show that they coincide anyway.

To finish this section we investigate the relations between the various measures:
\begin{lemma}[relations between measures]\label{lemma:RBM} For the measures defined above we have the following relations:
\begin{enumerate}
\item {$\mu^{d_D}=\mu^{F(D)}$} \label{enum:mudD=muFD}
\item {$\mu^{\mathcal{N}}$} and {$\mu^{F(D)}$} are equivalent, i.e.\ they have the same null-sets (and atoms). \label{enum:muNemuFD}
\item $\mu_f^D$ is continuous with respect to $\mu^{\mathcal N}$: $\mu_f^D(B)\leq \|f\|^2\mu^{\mathcal N}(B)$ \label{enum:mufDcmuN}
\end{enumerate}
\end{lemma}
\begin{proof}
\begin{enumerate}
\item {$\mu^{d_D}=\mu^{F(D)}$} just by definition, since $d_D(E,\chi)=0$ if and only if $E\in\spec D_\chi$.   
\item For every Borel set $B\subset\R$ we have:
\begin{align*}
\mu^{\mathcal{N}}(B) &= \int_{\hat\Gamma}\tr P_B^{D_\chi}\,d\chi \\
\intertext{and}
\tr  P_B^{D_\chi} \ne 0 &\Leftrightarrow P_B^{D_\chi} \ne 0 \\
&\Leftrightarrow B\cap \spec D_\chi\ne \emptyset \\
&\Leftrightarrow \exists{E\in B}: d_D(E,\chi)\ne0
\end{align*}
\item 
\begin{align*} \mu_{f_\chi}^{D_\chi} &= \<f_\chi|P_B^{D_\chi}f_\chi> \\
 &\leq \|f_\chi\|^2 \tr P_B^{D_\chi}
\end{align*}
\qed
\end{enumerate}
\end{proof}
%%%%%%%%%%%%%%%%%%%%%%%%%%%%%%%
\section{Spectral consequences}\label{sec:SC}
So far the operators $D_\chi$ act on different bundles, therefore \emph{a priori} on different domains.
We want to construct an isomorphism  $S_\chi:F_{\chi_0}\rightarrow F_\chi$ of Hermitian line bundles such that  $D_\chi':=(\ONE_{E'}\otimes S_\chi^*)\bar D_\chi (\ONE_{E'}\otimes S_\chi)$ becomes a real-analytic family of operators on  $E_{\chi_0}$.
We get the real-analytic structure on $\hat\Gamma$ by choosing an isomorphism $\hat\Gamma\simeq T^n\times\Sigma$, where $\Sigma$ is the (discrete) torsion part. 
Then the real-analytic standard structure of the  $n$-torus. 
(As already mentioned,  $\Gamma$ is finitely generated, so that indeed $n<\infty$.)
$T^n$ induces a real-analytic structure on $\hat\Gamma$.

Note that in our situation the sequence of coverings
\[ \begin{CD} \tilde X @. &\xrightarrow{\pi_1(X)} X \xrightarrow{\Gamma}& @. M\\
               \|      @. @. @. \| \\
              \tilde M @. @>\quad\pi_1(M)\quad>> @. M 
\end{CD}\]
induces the exact sequences
\[ \begin{CD}
0 @>>> \pi_1(X) @>>> \pi_1(M) @>>> \Gamma @>>> 0 \\
0 @<<< \widehat{\pi_1(X)} @<<< \widehat{\pi_1(M)} @<<< \hat\Gamma @<<< 0 
\end{CD} \]
of groups. Here, $\hat G$ defines the character group of a locally compact topological group $G$.

So we can embed the characters from $\hat\Gamma$ (injectively) into $\widehat{\pi_1(M)}$. 
This allows to consider the bundle $F_\chi=X\times_\chi\C$ as a  bundle $\tilde M\times_\chi\C$ associated to $\tilde M$:

\begin{proposition}[real-analytic family]\label{prop:RAF}
Let $\chi_0\in\hat\Gamma$. 
Then for every $\chi$ in the connected component of $\chi_0$ there exists an isomorphism $S_\chi:F_{\chi_0}\rightarrow F_\chi$ of Hermitian line bundles.
If we define $D_\chi':=(\ONE_{E'}\otimes S_\chi^*)\bar D_\chi (\ONE_{E'}\otimes S_\chi)$ then $(D'_\chi)_{\chi\in\Gamma}$ becomes a real-analytic family with constant domain of definition (family of type A).
\end{proposition}
\begin{proof}
First assume $\chi_0=1$. Let $\tilde x_0\in\tilde M$ be an arbitraily choosen base point. 
For $\omega\in\Omega^1(M),d\omega=0,$ denote by $\tilde\omega\in\Omega^1(\tilde M)$ the pull back to the covering space. Then
\[ \tilde s_\omega(\tilde x):=\exp\left(\imath\int_{\tilde x_0}^{\tilde x}\tilde\omega\right) \]
is well-defined since $d\tilde\omega=0$ and $H_1(\tilde M,\Z)=0$. For $\gamma\in\pi_1(M)$ one has
\begin{align*}
\tilde s_\omega(\gamma\tilde x) &= \exp\left(\imath\int_{\tilde x_0}^{\tilde x}\tilde\omega+\imath\int_{\tilde x}^{\gamma\tilde x}\tilde\omega\right) \\
 &= \tilde s_\omega \exp\left(\int_{c(\gamma)}\omega\right) \\
 &= \tilde s_\omega(\tilde x)\chi_\omega(\gamma)
\end{align*}
with the character $\chi_\omega\in\widehat{\pi_1(M)}$, defined by 
\[ \chi_\omega(\gamma):=\exp\left(2\pi\imath\int_{c(\gamma)}\omega\right) \]
for some choice of a closed path $c(\gamma)$ representing the class $\gamma\in\pi_1(M)$.
This is well-defined because $\omega$ is closed. On the other hand it is easily seen \cite[see e.g.][]{KatSun:HCGCRS,Gru:BTQMS} that every
 character in $\left(\widehat{\pi_1(M)}\right)_0$ may be represented in the form $\chi_\omega$ for some $\omega$. 
Especially this is possible for every character in $(\hat\Gamma)_0$.

Thus $\tilde s_\omega$  defines a section of $s_\omega\in C^\infty(\tilde M\times_{\chi_\omega}\C)$ with pointwise norm 1 and hence an Hermitian line bundle isomorphism
\begin{align*}
S_{\chi_\omega}: M\times\C&\rightarrow \tilde M\times_{\chi_\omega}\C, \\
 f&\mapsto fs_\omega \\
\intertext{so that}
\ONE_{E'}\otimes S_\chi: E'&\rightarrow E_{\chi_\omega} .
\end{align*}

Finally $d\tilde s_\omega(\gamma\tilde x)=\imath\tilde s_\omega\tilde\omega$. 
Therefore we have on $C^\infty_0(M)$
\begin{align*}
D_{\chi_\omega}' &:= (\ONE_{E'}\otimes S_{\chi_\omega}^*)D_{\chi_\omega} (\ONE_{E'}\otimes S{\chi_\omega}) \\
 &= D + T_{\chi_\omega},
\end{align*}
where $T_{\chi_\omega}$ is a differential operator of order $\ord(D)-1$ with smooth bounded coefficients; these are real-analytic in $\chi$ because they contain $\omega$ only polynomially. 
So the domain of $D_\chi'$ is the Sobolev space $W^{\ord(D)}(E')$, independent of $\chi$.

If $\chi_0\neq1$ is in an arbitrary connected component of $\hat\Gamma$ we can use $E_{\chi_0}\simeq E_{\chi_0}\otimes(M\times\C)$ and get an Hermitian line bundle isomorphism
\[ \ONE_{E_{\chi_0}}\otimes S_{\chi_\omega}: E_{\chi_0}\rightarrow E_{\chi_0\chi_\omega} \]
with the same properties as above. 
So we have the desired real-analytic family over every connected component.
\qed \end{proof}

\begin{lemma}[$\zeta$-regularized determinant]\label{lemma:ZRD}
If $D$ is bounded-below then every fiber $\bar D_\chi'-\lambda$ has a $\zeta$-regularized determinant 
\[ d_D(\lambda,\chi) = \det\nolimits^\zeta (\overline{D_\chi'}-\lambda) \]
such that $d_D(\lambda,\chi)=0\Leftrightarrow \lambda\in\spec\bar D_\chi'$. $d_D$ is real-analytic in $\chi$ and analytic in $\lambda$.
\end{lemma}
\begin{proof}
$\bar D_\chi'$ is bounded below ($\bar D\geq C\Rightarrow\bar D_\chi\geq C$ because of \eqref{equ:Dchi=D|Echi}),
 self-adjoint, elliptic, and defined on a compact manifold.
Therefore it has a $\zeta$-regularized determinant $d_D(\lambda,\chi)=\det\nolimits^\zeta (\overline{D_\chi'}-\lambda)$, which is analytic in $\lambda$ and fulfills $d_D(\lambda,\chi)=0\Leftrightarrow \lambda\in\spec\bar D_\chi'$  \cite[see, e.g.,][]{Gil:ITHEASIT}. 
Since the family $(D'_\chi)_{\chi\in\hat\Gamma}$ depends real-analytically on $\chi$ the coefficients in the heat trace asymptotics do so.
The asymptotics is uniform in $\chi$, so the  $\zeta$-function as well as the $\zeta$-regularized determinant depend real-analytically on $\chi$.
\qed \end{proof}

From \citet{HemHer:BGPMH} we cite
\begin{lemma}[real analyticity and measure]\label{lemma:RAM}
\begin{enumerate}
\item Let $O\subset\R^n$ be open and connected, $h:O\rightarrow\C$ real-analytic. 
If $h^{-1}(\{0\})$ has zero Lebesgue measure then $h\equiv0$. \label{enum:h=0}
\item Let $O\subset\R^n$ be open and connected, $I\subset\R$ an open intervall, $f:I\times O\rightarrow \C$ real-analytic, $f\not\equiv0$. 
For Borel sets $B\subset\R$ 
\begin{equation}
\mu^f(B)=\meas\{k\in O\mid \exists\lambda\in B:f(\lambda,k)=0\} 
\end{equation}
defines a quasi-measure. Then we have 
\begin{enumerate}
\item $\mu^f$ has no singular continuous component (i.e.\ there is no decomposition containing a singular continuous component). \label{enum:NSCC}
\item The atoms of $\mu^f$ are discrete in $\R$. \label{enum:AD}
\end{enumerate}
\end{enumerate}
\end{lemma}

\begin{theorem}[spectral nature of abelian-periodic operators]\label{theorem:SNAPO}
Let $D$ be a symmetric elliptic abelian-periodic differential operator. 
Then we have for the spectrum of $\bar D$:
\begin{enumerate}
\item The singular continuous spectrum is empty. \label{enum:singleer}
\item The density of states measure $\mu^{\mathcal N}$ has no singular continuous component.\label{enum:DOSnotsing}
\item The point spectrum is discrete as subset of $\R$. \label{enum:Punktspektrum diskret}
\item If $\lambda$ is an eigenvalue of $\bar D$ then there exists a connected component $\Lambda$ of $\hat\Gamma$ such that for $\chi\in\Lambda$: $\lambda\in\spec \overline{D_\chi}$.  \label{enum:EW auf Komponente}
\end{enumerate}
\end{theorem}
\begin{proof}
First we show that we may assume $\bar D\geq0$ without loss of generality.
For this we note that the spectral projections $P^D$ and $P^{D^2}$ of $\bar D$ and $\bar D^2$ are related by:
\begin{align}
P^{D^2}(\lambda) &= 1_{(-\infty,\lambda]}\left(\bar D^2\right) \notag\\
 &= \left(1_{(-\infty,\lambda]}\circ (\cdot)^2\right)\left(\bar D\right) \notag\\
 &= 1_{\left[-\sqrt[\sgn]{\lambda},\sqrt[\sgn]{\lambda}\right]}\left(\bar D\right) \notag\\
 &= P^D_{\left[-\sqrt[\sgn]{\lambda},\sqrt[\sgn]{\lambda}\right]} \notag \\
 &= \begin{cases}P^D\left(\sqrt\lambda\right)-P^D\left(-\sqrt\lambda\right),&\text{ if }\lambda\geq0,\\
    0&\text{ else,}\end{cases} \label{equ:PD2PD} 
\end{align}
where we set $\sqrt[\sgn]{\lambda}:=\sgn\lambda\sqrt{|\lambda|}$. 
For the point spectrum as well as the whole spectrum as a set we can use the spectral mapping theorem: 
\[ \spec_p \left(\bar D^2\right) = \left(\spec_p \bar D\right)^2,\quad \spec \left(\bar D^2\right) = \left(\spec \bar D\right)^2 \]
Because of the continuity of $(\cdot)^2$ and $\sqrt\cdot$ this implies $\spec_{p.p.} \left(\bar D^2\right) = \left(\spec_{p.p.} \bar D\right)^2$, da $\spec_{p.p.}=\overline{\spec_p}$.
For \ref{enum:Punktspektrum diskret} and \ref{enum:EW auf Komponente} it therefore suffices to consider $\bar D^2$.

If $\bar D^2$ has no singular continuous spectrum then, given a set $B$ with zero Lebesgue measure and $B\subset\R\setminus\spec_{p.p.} \left(\bar D^2\right)$ and an arbitrary $s\in{\cal D}\left(\bar D^2\right)\setminus0\subset{\cal D}\left(\bar D\right)\setminus0$, the spectral measure $\mu_{s,s}^{D^2}(B)$ at $s$ vanishes on $B$. 
If we put
\begin{align*} B&:=\left\{\lambda\in\R\mid\lambda^2\in \tilde B\right\}=\sqrt{|\tilde B|}\cup\left(-\sqrt{|\tilde B|}\right), \\
\Tilde{\Tilde{B}}&:= \left\{\lambda\in\R\mid|\lambda|\in \tilde B\right\}=B^2\cup\left(-B^2\right)=\tilde B\cup\left(-\tilde B\right)\supset \tilde B
\end{align*}
for a Lebesgue zero-set $\tilde B\subset\R\setminus\spec_{p.p.}\bar D$ then $B$ has Lebesgue measure zero and contains, by the spectral mapping theorem, no elements of $\spec_{p.p.} \left(\bar D^2\right)$. 
From~\eqref{equ:PD2PD} we get
{\allowdisplaybreaks \begin{align*}
0&=\mu_{s,s}^{D^2}(B) \\
&= \int_\R 1_B(\lambda)\,d\<s|P^{D^2}(\lambda)s> \\
&=\int_{\R_+} 1_B(\lambda)\,d\<s|P^D\left(\sqrt\lambda\right)s>-\int_{\R_+} 1_B(\lambda)\,d\<s|P^D\left(-\sqrt\lambda\right)s>\\
&=\int_{\R_+} 1_B\left(\lambda^2\right)\,d\<s|P^D(\lambda)s>-\int_{\R_+} 1_B(\lambda^2)\,d\<s|P^D(-\lambda)s>\\
&=\int_{\R} 1_B\left(\lambda^2\right)\,d\<s|P^D(\lambda)s>\\
&=\mu_{s,s}^D\left(\Tilde{\Tilde{B}}\right) \\
&\geq\mu_{s,s}^D\left(\tilde{B}\right).
\end{align*}}
Since ${\cal D}\left(\bar D^2\right)$ is dense one has $\spec_{s.c.}\bar D=\emptyset$.
So, even for \ref{enum:singleer} it is sufficient to consider $\bar D^2$.

Finally, by assumption $D^2$ is a symmetric elliptic abelian-periodic differential operator.
Given  $\overline{D^2}=\bar D^2\geq 0$ because of the essential self-adjointness we can assume $D\geq0$ in the following.

\bigskip 
Let $d_D$ be the real-analytic family of $\zeta$-determinants of $D_\chi$ which exists by lemma~\ref{lemma:RAM}.
\begin{enumerate}
\item[\ref{enum:EW auf Komponente}.] If $\lambda\in\spec_p(\bar D)$ then, by the general theory for direct integrals,
\begin{align*} 0&< \meas(\{\chi\in\hat\Gamma\mid\lambda\in\spec_p(\bar D_\chi \}) \\
&= \meas(\{\chi\in\hat\Gamma\mid d_D(\lambda,\chi)=0 \}) \\
&= \meas(h^{-1}(\{0\}))\text{ with} \\
h&=d_D(\lambda,\cdot).
\end{align*}
Since $\hat\Gamma$ has only finitely many  connected components there must be a component $\Lambda$ with $\meas(h^{-1}({0})\cap\Lambda)>0$ . 
Then $h|_\Lambda$ fulfills the assumptions of lemma~\ref{lemma:RAM}.\ref{enum:h=0} so that  $h|_\Lambda\equiv0$ and thus $\lambda\in\spec \bar D_\chi$ for $\chi\in\Lambda$.
\item[\ref{enum:Punktspektrum diskret}.] As above
\begin{align*}
\lambda\in\spec_p \bar D&\Leftrightarrow \meas(\{\chi\in\hat\Gamma\mid\lambda\in\spec_p(\bar D_\chi \}) >0 \\
&\Leftrightarrow \meas(\{\chi\in\hat\Gamma\mid d_D(\lambda,\chi)=0 \}) >0 \\
&\Leftrightarrow \mu^{d_D}(\{\lambda\})>0\\
&\Leftrightarrow \lambda \text{ is an atom of } \mu^{d_D}.
\end{align*}
But by lemma~\ref{lemma:RAM}.\ref{enum:AD} the atoms are discrete.
\item[\ref{enum:singleer}.] 
Consider  a Borel set $B\subset\R$ with $B\cap\spec_p D=\emptyset$ such that $B$ contains no atoms of $\mu^{d_D}$. 
If $\meas(B)=0$ then  $\mu^{d_D}(B)=0$ since $\mu^{d_D}$ has, by lemma~\ref{lemma:RAM}, \ref{enum:NSCC}, no singular continuous component with respect to the Lebesgue measure $\meas$.
Since $\mu^{\mathcal N}$ and $\mu^{d_D}$ are equivalent (by lemma~\ref{lemma:RBM}, \ref{enum:mudD=muFD}. and \ref{enum:muNemuFD}.) we have $\mu^{\mathcal N}(B)=0$. Since $\mu_f^D$ is continuous with respect to $\mu^{\mathcal N}$ (by lemma~\ref{lemma:RBM}, \ref{enum:mufDcmuN}.) we conclude $\mu_f^D(B)=0$.
Thus $\mu_f^D$ has no singular continuous component with respect to $\meas$.

\item[\ref{enum:DOSnotsing}.] From \ref{enum:singleer}. we get the assertion about the density of states, since $\mu^{\mathcal N}$ and $\mu^{F(D)}$ are equivalent by lemma~\ref{lemma:RBM}.\ref{enum:muNemuFD}.
\qed \end{enumerate}
\end{proof}

\begin{remark}
For the periodic Schr\"odinger operator one has absolutely continuous spectrum only.
On the other hand, the Landau operator (constant magnetic field, dimension 2) has only pure point spectrum. 
This shows that the theorem is optimal with respect to possible  measure types under the given conditions.
\end{remark}

\begin{remark}
As is well known, any possibly occuring eigenvalue automatically has infinite degeneracy (from periodicity under the infinite group). So the spectrum is purely essential.
\end{remark}

%% file: main.bbl
%auto-ignore
\ifx\undefined\allcaps\def\allcaps#1{#1}\fi
  \ifx\undefined\nop\newcommand{\nop}[1]{}\fi
  \ifx\undefined\single\newcommand{\single}[1]{#1}\fi
  \ifx\undefined\SwapArgs\newcommand{\SwapArgs}[2]{#2#1}\fi
  \ifx\undefined\translationof\newcommand{\translationof}{English Translation
  of }\fi \ifx\undefined\submitted\newcommand{\submitted}{Submitted }\fi
  \ifx\undefined\submittedto\newcommand{\submittedto}{Submitted to }\fi
  \ifx\undefined\privcomm\newcommand{\privcomm}{Private communication}\fi
  \providecommand{\inpreparation}{In preparation}
  \providecommand{\toappearin}{To appear in }